\documentclass[11pt,a4paper]{article}
\usepackage{jheppub}
\usepackage{setspace,braket}
\usepackage{amsmath, amssymb, bm}
\usepackage{etoolbox}

    \makeatletter
    \patchcmd{\maketitle}{\@fpheader}{}{}{}
\makeatother
\def\be{\begin{equation}}
\def\ee{\end{equation}}

\def\({\left(}
\def\){\right)}
\def\[{\left[}
\def\]{\right]}

\newcommand{\bea}{\begin{eqnarray}}
\newcommand{\eea}{\end{eqnarray}}

\def\d#1#2{\frac{\displaystyle #1}{\displaystyle #2}}
\def\no{\nonumber}

\numberwithin{equation}{section}
\renewcommand{\thesection}{\arabic{section}}

\begin{document}
\renewcommand{\thefootnote}{\fnsymbol{footnote}}

\title{Peculiar $P-V$ criticality of topological Ho\v{r}ava-Lifshitz black holes}
\author[a]{Meng-Sen Ma,}
\author[b]{Rui-Hong Wang}
\affiliation[a]{Institute of Theoretical Physics, Shanxi Datong
University, Datong 037009, China}
\affiliation[b]{College of Information Science and Technology, Hebei Agricultural University, Baoding 071001, China}
\emailAdd{mengsenma@gmail.com; ms\_ma@sxdtdx.edu.cn}

\abstract{

We demonstrate the existence of $P-V$ criticality of the topological Ho\v{r}ava-Lifshitz(HL) black holes with a spherical horizon $(k=1)$ in the extended phase space.
With the electric charge, we find that the critical behaviors of the HL black hole are nearly the same as those of van der Waals(VdW) system.
For the uncharged case, the HL black hole has a peculiar $P-V$ criticality. The critical behavior is completely controlled by a parameter $\epsilon$, but not the temperature $T$.
When $\epsilon$ is larger than a critical value $\epsilon_c$, no matter what the temperature is, there will be the first-order phase transition. Moreover, we find that there is an infinite number of critical points which form a `` critical curve ". As far as we know, this is the first time to find this kind of peculiar $P-V$ criticality.

}
\maketitle
\onehalfspace

%%%%%%%%%%%%%%%%%%%%%%%%%%%%%%%%%%%%%%%
%%%%%%%%%%%%%%%%%%%%%%%%%%%%%%%%%%%%%%%
%%%%%%%%%INTRODUCTION%%%%%%%%%%%%%%%%%%%
%%%%%%%%%%%%%%%%%%%%%%%%%%%%%%%%%%%%%%%
\renewcommand{\thefootnote}{\arabic{footnote}}
\setcounter{footnote}{0}
\section{Introduction}
\label{intro}

Since the works of Hawking and Bekenstein\cite{Bekenstein-1973,Hawking-1975}, thermodynamic properties of gravitational systems, especially black holes, have been extensively studied.
Until now, there have been many approaches to calculating the temperature and entropy of black holes. Besides, the phase structure and critical phenomena of black holes have also been studied for quite a long time\cite{Hut.1977,Davies.1978,Sokolowski.1980,Pavon.1991,Lau.1994,Carlip.2003,Lundgren.2008}. Since the pioneering work of Hawking and Page\cite{Hawking.1983} and the popularity of AdS/CFT correspondence, phase transitions of black holes in AdS space have aroused much interest\cite{Chamblin.1999,Chamblin.1999b,Cvetic.1999,Cvetic.1999b,Peca.1999,Wu.2000,Biswas.2004,Myung.2005,Dey.2007,Myung.2008,Quevedo.2008,Cadoni.2010,Liu.2010,Sahay.2010,Banerjee.2011,Ma.2014,Ma.2014b}.

Recently, physicists reconsidered the critical behaviors of black holes in extended phase space, where the cosmological constant is treated as the thermodynamic pressure and its conjugate quantity as the thermodynamic volume of the black holes. It is shown that the $P-V$ criticality of some black holes in this extended phase space are very similar to those of a van der Waals (VdW) liquid/gas system\cite{Dolan.2011,Kubiznak.2012,Cai.2013,Chen.2013,Hendi.2013,Altamirano-2014,Mo.2014,Xu.2014,Ma.2015,Ma.2015b,ZhaoHH.2015}. Further, in this extended phase space other interesting critical behaviors such as reentrant phase transition, triple critical point, isolated critical point for several black holes have been explored\cite{Altamirano.2013,Wei.2014,Dolan.2014,Frassino.2014}. Thus, we want to know whether there are still other peculiar critical phenomena in the extended phase space. This is the first motivation of the present work.

 Ho\v{r}ava-Lifshitz (HL) gravity, which was proposed by Ho\v{r}ava, is a power-counting renormalizable gravity theory and can be regarded as an ultraviolet complete candidate for general relativity\cite{Horava.2009}. The HL gravity is constructed under two conditions: projectability and detailed balance. The breaking of projectability can simplify the theory.
 One has more freedom to construct the invariants under foliation of spacetime-preserving diffeomorphism. With the detailed balance, it is shown that the HL gravity has some problems, such as parity violating, ultraviolet instability and strong coupling problem\cite{Calcagni.2009,Charmousis.2009,Sotiriou.2009,Calcagni.2010}. Therefore, it is also reasonable to abandon the detailed balance.  Relaxing the projectability condition and deviating from detailed balance, some spherically symmetric black hole solutions has been derived in \cite{Lu.2009,Cai.2009c}.  The thermodynamic quantities of some HL black holes are calculated in \cite{Cai.2009b,Myung.2009,Kiritsis.2010,Liu.2014}. In \cite{Cao.2011} the authors studied the phase transition of a charged topological HL black hole with general dynamical parameter $\lambda$ and the phase transition of a Kehagias-Sfetsos black hole. In \cite{Majhi.2012}, the phase transition and scaling behavior of charged HL black holes with $\lambda=1$ were studied. Through thermodynamic metrics, in \cite{Mo.2013} the authors analyzed a unified phase transition of the charged topological HL black hole in the case of $\lambda=1$. And the same author concluded that no $P-V$ criticality exists in a topological charged HL black hole\cite{Mo.2015}. In the framework of horizon thermodynamics proposed by Padmanabhan\cite{Padmanabhan.2002}, we also studied the HL black hole and found a universal phase structure and the $P-V$ criticality\cite{Ma.2017}.

 In this paper, we would study the $P-V$ criticality of topological HL black holes in the extended phase space under the more general consideration of nonzero $\epsilon$ (violation of the detailed-balance condition). We want to clarify the influence of the parameter $\epsilon$ on the thermodynamic properties and critical behaviors of the HL black holes. This is the second motivation of the present work. In the uncharged case, we find a peculiar $P-V$ criticality, for which there is an infinite number of critical points, in fact, a ``critical curve". For the charged HL black hole, there is the usual $P-V$ criticality similar to that of a VdW system. The parameter $\epsilon$ plays an important role in the generation of these critical phenomena.

The plan of this paper is as follows:
In Sec.2 we introduce the HL black hole and the thermodynamic quantities.
We also present the necessary demonstrations on some notations.
In Sec.3 we analyze the conditions under which the temperature and the entropy of the HL black hole are positive.
In Sec.4 we study the $P-V/r_{+}$ criticality of the HL black hole in the uncharged and charged case, respectively.
In Sec.5 we summarize our results and discuss the possible future directions.

\section{Thermodynamic quantities of Ho\v{r}ava-Lifshitz black hole}

In this section, we simply introduce the Ho\v{r}ava-Lifshitz black hole and its thermodynamic quantities studied in \cite{Lu.2009,Cai.2009c}.
The action of HL gravity without the detailed-balance condition is

\begin{eqnarray}
\label{action}
I &=& \int dtd^3x ({\cal L}_0 +(1-\epsilon^2){\cal L}_1 +{\cal L}_m),\nonumber \\
 {\cal L}_0 &=& \sqrt{g}N \left \{\frac{2}{\kappa^2}
(K_{ij}K^{ij}-\lambda K^2) +\frac{\kappa^2\mu^2 (\Lambda
R-3\Lambda^2)}{8(1-3\lambda)}\right \},  \nonumber \\
 {\cal L}_1  &=& \sqrt{g}N \left \{\frac{\kappa^2\mu^2(1-4\lambda)}{32(1-3\lambda)}R^2
-\frac{\kappa^2}{2\omega^4}Z_{ij}Z^{ij} \right\},
\end{eqnarray}
where $Z_{ij}=\left(C_{ij}-\frac{\mu
\omega^2}{2}R_{ij}\right)$ with $C_{ij}$ the Cotton tensor. In this theory, there are several parameters: $\epsilon$, $\kappa^2$, $\lambda$, $\mu$, $\omega$ and $\Lambda$. ${\cal L}_m$ stands for the Lagrangian of other matter fields.

Compared with general relativity, there will be the relations for the parameters:
\be\label{constants}
c=\frac{\kappa^2\mu}{4}\sqrt{\frac{\Lambda}{1-3\lambda}}, \ \
G=\frac{\kappa^2 c}{32\pi}, \ \
\tilde\Lambda=\frac{3}{2}\Lambda,
\ee
where $G,~c$ and $\tilde\Lambda$ are Newton's constant, the speed of light, and the cosmological constant respectively. In the following work, we will take the natural units: $G=c=\hbar=k_B=1$,
so  $\kappa^2= 32 \pi$ and $\mu^2 = -1/( 32 \pi^2\Lambda)$.

Because only for the case with $\lambda=1$, general relativity can be recovered in the large distance approximation, we only consider $\lambda=1$ in the following.
In this case, from Eq.(\ref{constants}), one can see that $\Lambda$ must be negative.

$\epsilon=0$ corresponds to the so-called detailed-balance condition, under which HL gravity turns out to
be intimately related to topological gravity in three dimensions and the geometry of the Cotton tensor.
For the case with $\epsilon=1$, HL gravity returns back to general relativity and the HL black hole becomes a Schwarzschild-(A)dS black hole. Therefore, we will consider the general values of $\epsilon$ in the region $0\leq \epsilon^2 \leq 1$ below.

For a static, spherically symmetric black hole, the metric ansatz can be written as
\be\label{staticmetric}
ds^2 =-\tilde N^2(r)f(r) dt^2 +\frac{dr^2}{f(r)} +r^2
d\Omega_k^2,
\ee
where $d\Omega_k^2$ denotes the line element for a two-dimensional
Einstein space with constant scalar curvature $2k$. Without loss of generality, one can take $k=1$ (spherical/elliptic
horizons), $k = 0$ (flat horizons), and $k=-1$ (hyperbolic horizons).

When a Maxwell field exists, it is shown that the metric function $\tilde N(r)=1$ and  $f(r)$ is given by\cite{Cai.2009c}
\be\label{metric}
f(r)=k+\frac{x^2}{1-\epsilon ^2}-\frac{\sqrt{\left(1-\epsilon ^2\right) \left(m x-q^2/2\right)+x^4 \epsilon ^2}}{1-\epsilon ^2},
\ee
where $x=\sqrt{-\Lambda}r$ and  $m$ and $q$ are integration constants. They are related to the black hole mass and electric charge:
\be\label{mq}
M=\frac{\kappa ^2 \mu ^2 \Omega _k \sqrt{-\Lambda }}{16}  m=\frac{m}{\sqrt{-\Lambda }}, \quad Q=\frac{\kappa ^2  \mu ^2 \Omega _k\sqrt{-\Lambda }}{16}  q=\frac{q}{\sqrt{-\Lambda }},
\ee
where we have taken the natural units and set $\Omega _k=16\pi$. The event horizon is defined through $f(x_{+})=0$, where $x_{+}$ denotes the largest root of $f(x)$ and its complete expression is given in the Appendix.

If we consider the cosmological constant $\tilde\Lambda$ as a variable and identify it with pressure $P=-\frac{\tilde\Lambda}{8\pi}$, then according to Eq.(\ref{constants}), there is $\Lambda=-\frac{16\pi P}{3}$.

From Eqs.(\ref{metric}) and (\ref{mq}), the black hole mass is
\be
M=\frac{-9 k^2 \left(\epsilon ^2-1\right)+96 \pi  k P r_+^2+256 \pi ^2 P^2 r_+^4+24 \pi  P Q^2}{48 \pi  P r_+},
\ee
where $r_{+}$ is the position of the event horizon of the HL black hole.

According to the metric function, one can easily derive the temperature:
\be\label{tem}
T=\frac{3 k^2 \left(\epsilon ^2-1\right)+32 \pi  P r_+^2 \left(k+8 \pi  P r_+^2\right)-8 \pi  P Q^2}{8 \pi  r_+ \left(-3 k \epsilon ^2+3 k+16 \pi  P r_+^2\right)}.
\ee
The entropy of the charged topological HL black hole is
\bea\label{entropy}
S=\frac{8 \pi  P r_+^2-3 k \left(\epsilon ^2-1\right) \ln \left(4 \sqrt{\frac{\pi }{3}} \sqrt{P} r_+\right)}{2 P}+S_{0}.
\eea
Clearly, it is independent of the electric charge $Q$. The integration constant $S_0$ in the entropy
cannot be fixed by some physical considerations. To determine $S_0$, one has to invoke the quantum theory of gravity as argued in \cite{Cai.2009c}. For $k=0$ case, the area law of the black hole entropy is recovered if setting $S_0=0$. Thus, for simplicity, we always set $S_0=0$ below (including the cases of $k=\pm 1$).

Other thermodynamic quantities can also be derived from the extended first law of black hole thermodynamics:
\be
dM=TdS+\Phi dQ+VdP.
\ee
Here we do not consider the parameter $\epsilon$ as a thermodynamic variable. In the extended phase space, the black hole mass now should be considered as the enthalpy, $M=H$.
Due to the existence of the logarithmic term in the entropy, no Smarr-like relation for the HL black hole exists.

In this paper we are only concerned with the fixed charge ensemble. Thus, the Gibbs free energy is now defined by
\be
G=H-TS=M-TS,
\ee
by which we can analyze the global thermodynamic stability and phase transition of the HL black hole.

\section{Positivity of entropy and temperature of Ho\v{r}ava-Lifshitz black holes}

\begin{figure}[htp]
\center{
\includegraphics[width=7cm,keepaspectratio]{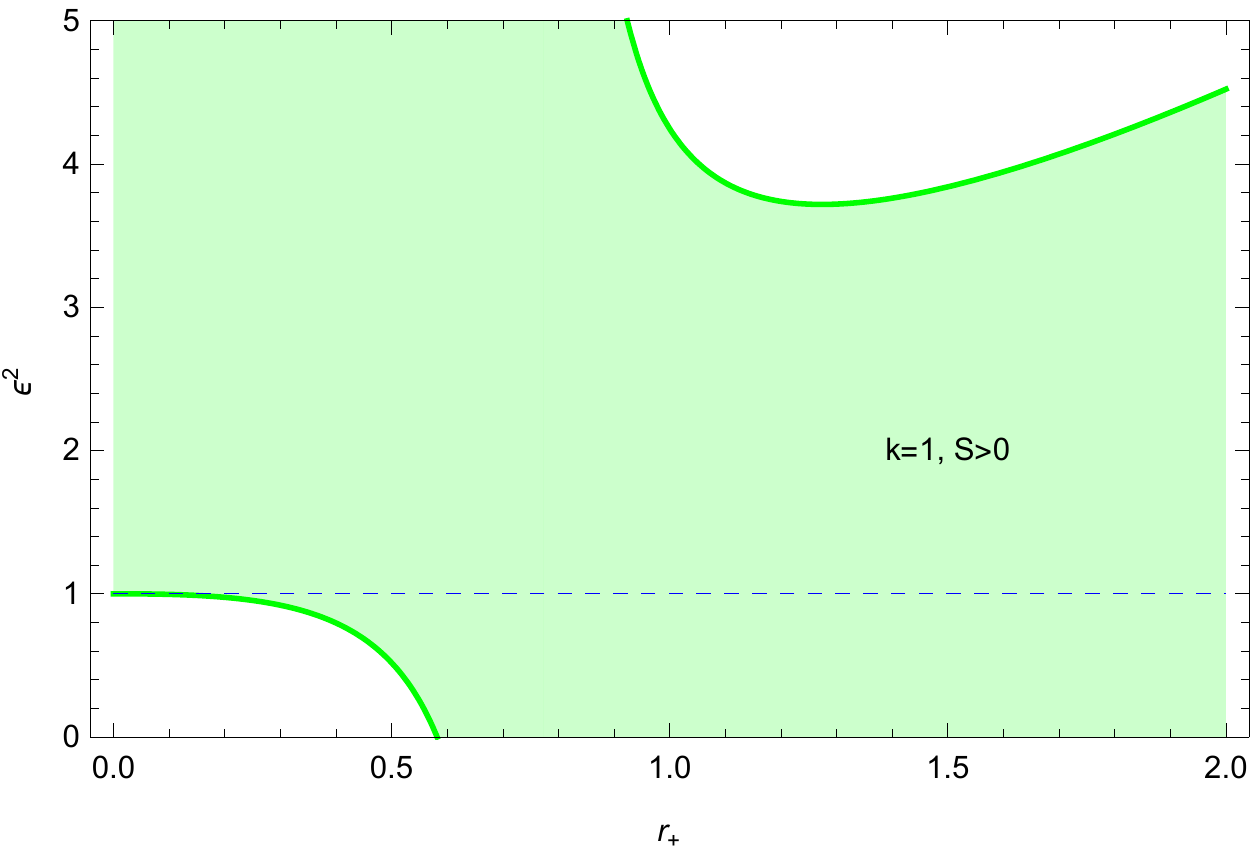}\hspace{0.2cm}
\includegraphics[width=7cm,keepaspectratio]{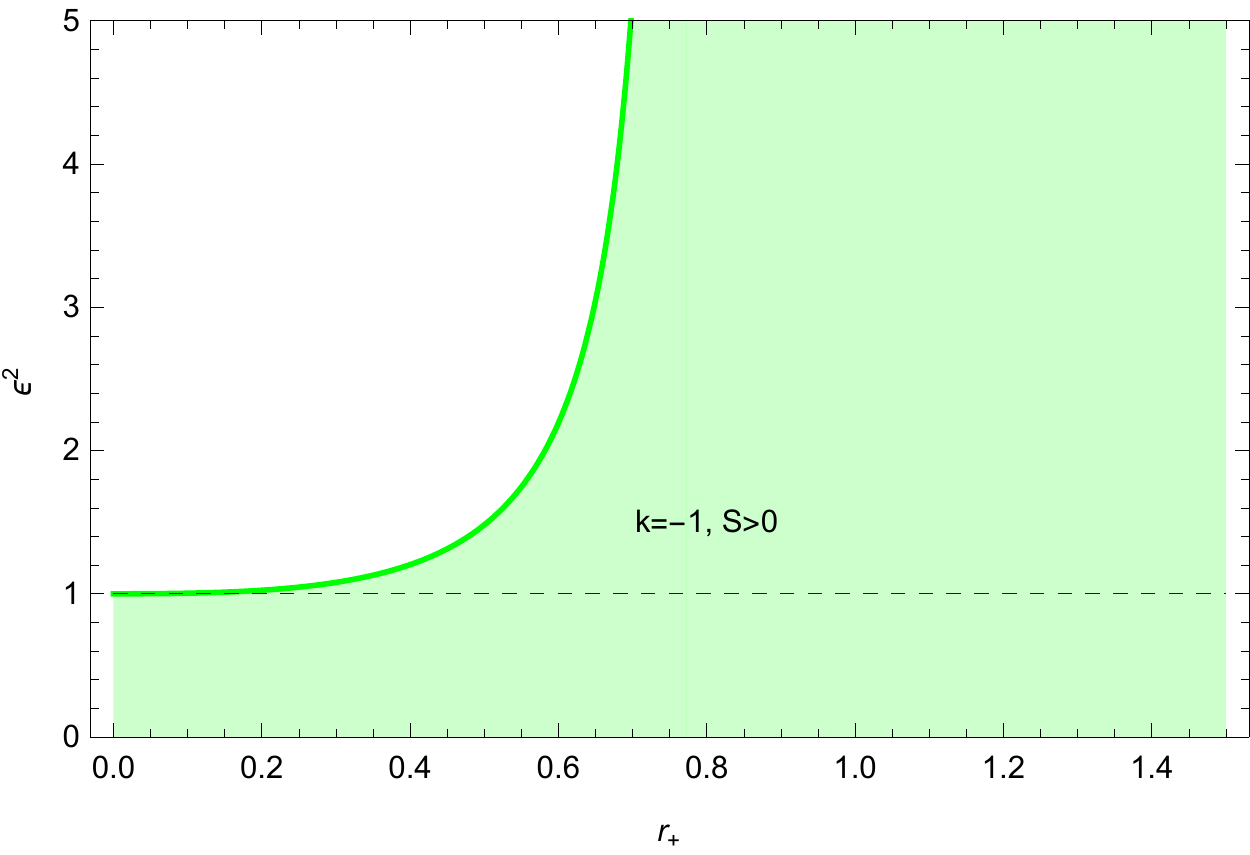}
\caption{The relations between $\epsilon^2$ and $r_{+}$ for the positive entropy. In the green shaded area, the entropy is positive. We have taken  $P=0.1$.}\label{pS}}
\end{figure}

In this section, we will analyze the conditions under which the entropy and the temperature of the topological HL black holes are positive. The pressure $P$ is clearly positive according to the definition.
According to Eq.(\ref{entropy}), the entropy is a function of $(P,~\epsilon, ~r_{+})$, and is independent of the electrical charge $Q$. After setting $S_0=0$, the entropy is not always positive except for the case $k=0$. To determine the positivity of $S$, we fix the pressure and analyze the relations between $\epsilon$ and $r_{+}$. In Fig.\ref{pS}, we show the relations between $\epsilon^2$ and $r_{+}$ for the positive entropy. We only pay attention to the case with $0\leq \epsilon^2\leq 1$. For $k=1$, it is shown that the scope of $r_{+}$ becomes larger and larger as $\epsilon^2$ increases, and $r_{+}$ cannot reach to zero until $\epsilon^2=1$. For $k=-1$, we find that the entropy is always positive so long as $\epsilon$ belongs to the region $0\leq \epsilon^2\leq 1$.

For the temperature, we only analyze the uncharged case $(Q=0)$ for simplicity. For the charged case, the analysis is similar. According to Eq.(\ref{tem}), we can also depict the $\epsilon^2-r_{+}$ diagrams for positive temperature.
For $k=1$, the lower branch corresponds to zero temperature and the upper branch corresponds to the divergent temperature. Only on the right-hand side of the two branches is the temperature positive. For $k=-1$, the lower branch
corresponds to the divergent temperature and the upper branch corresponds to the zero temperature. Outside the region enclosed by the two curves, the temperature is positive.

\begin{figure}[htp]
\center{
\includegraphics[width=7cm,keepaspectratio]{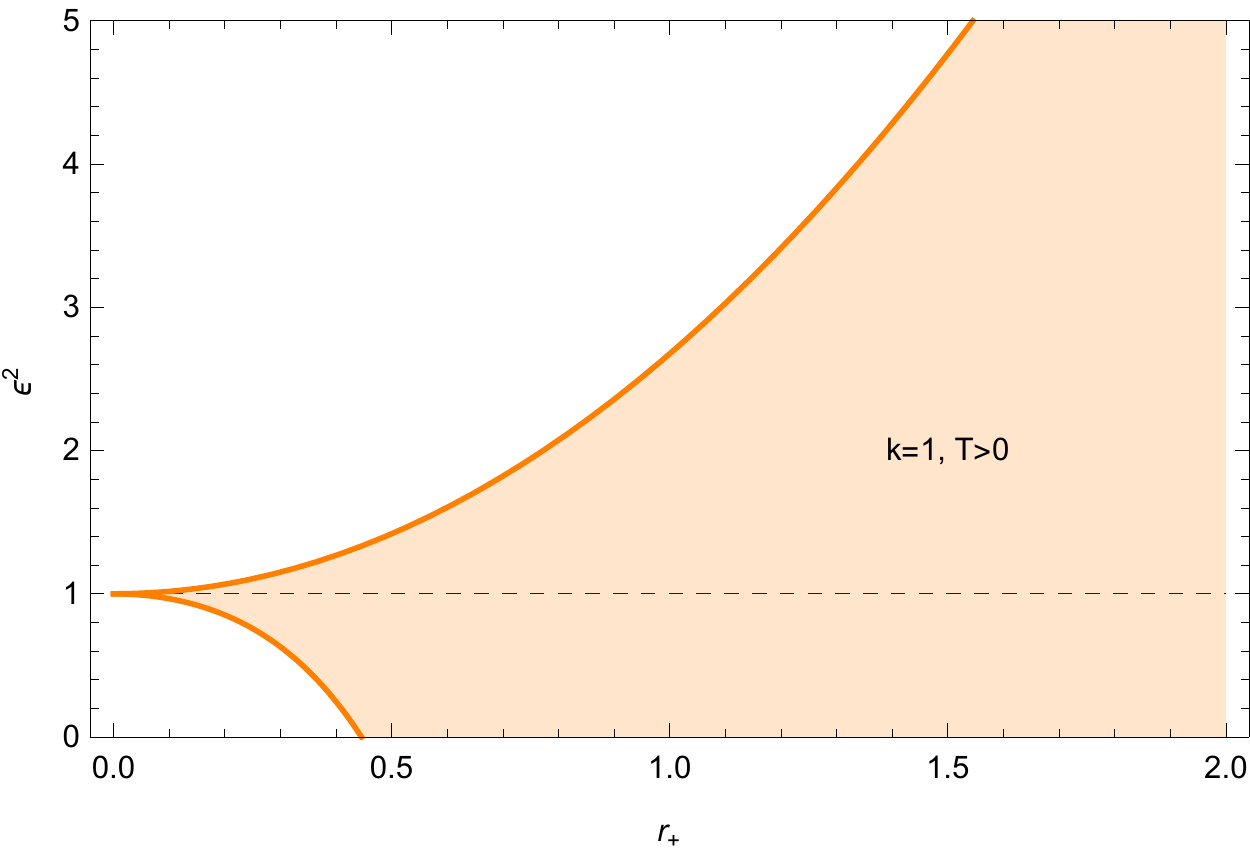}\hspace{0.2cm}
\includegraphics[width=7cm,keepaspectratio]{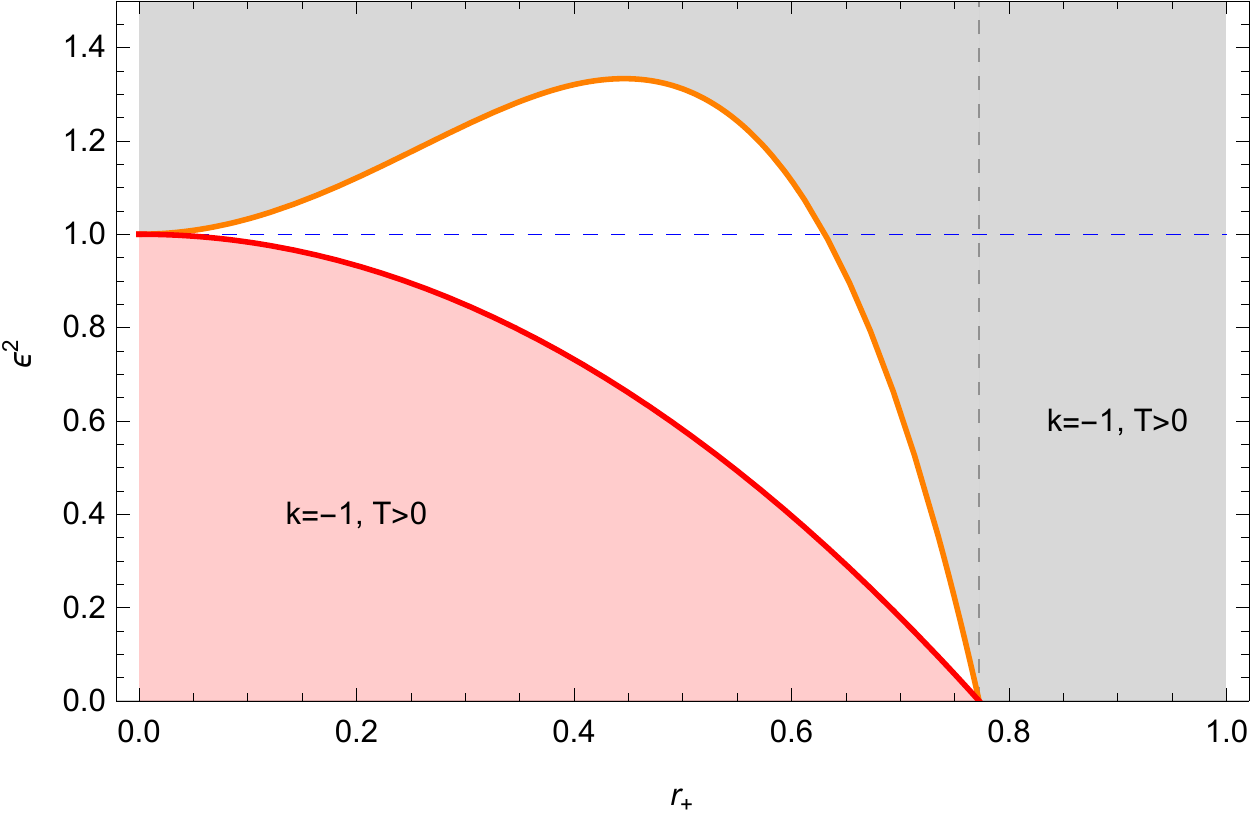}
\caption{The relations between $\epsilon^2$ and $r_{+}$ for the positive temperature. In the shaded area, the temperature is positive. We set $P=0.1$.}\label{pT}}
\end{figure}

The intersection parts of Figs.\ref{pS} and \ref{pT} are the regions where both the entropy and the temperature are positive. And the analysis in the next section should be restricted in these regions.

\section{$P-V$ criticality of Ho\v{r}ava-Lifshitz black hole}

According to Eq.(\ref{tem}), one can derive the pressure as a function of $(T,r_{+})$:
\bea\label{Pre}
P(T,r_{+})&=&\frac{\sqrt{16 k^2 r_+^4 \left(4-3 \epsilon ^2\right)-8 k Q^2 r_+^2-128 \pi  k r_+^5 T \left(3 \epsilon ^2-2\right)+Q^4+32 \pi  Q^2 r_+^3 T+256 \pi ^2 r_+^6 T^2}}{64 \pi  r_+^4}\no \\
&+&\frac{Q^2-4 k r_+^2+16 \pi  r_+^3 T}{64 \pi  r_+^4}.
\eea
After a series expansion, it has the form
\be
P=\frac{T}{2 r_+}+\frac{k-3 k \epsilon ^2}{16 \pi  r_+^2}-\frac{9 k^2 \epsilon ^2 \left(\epsilon ^2-1\right)}{128 r_+^3 \pi ^2 T}+O(r_{+}^{-4}).
\ee
By comparing the above equation with the van der Waals equation, one can easily find the specific volume $v \propto r_{+}$.
So we would not introduce the specific volume, but directly study the $P-r_{+}$ behavior.

If the HL black hole has the similar $P-V$ criticality to that of a VdW system, the critical points should satisfy the following equations:
\be
\frac{\partial P}{\partial r_{+}}=0, \quad \frac{\partial^2 P}{\partial r_{+}^2}=0.
\ee
It is shown that the two equations are too complicated to be directly solved. However, we can employ the implicit differentiation on Eq.(\ref{tem}) to obtain
\be
P'(r_{+})=\frac{-8 \pi  P }{A(r_{+},T,P)}\left[8 P r_+ \left(k-6 \pi  r_+ T\right)+3 k T \left(\epsilon ^2-1\right)+128 \pi  P^2 r_+^3\right],
\ee
with
\be
A(r_{+},T,P)=8 \pi  r_+ \left[32 \pi  r_+^3 P^2-3 k T \left(\epsilon ^2-1\right)\right]-3 k^2 \left(\epsilon ^2-1\right).
\ee
and
\be
P''(r_{+})=\frac{-64 \pi  P^2}{A(r_{+},T,P)} \left[k+12 \pi  r_+ \left(4 P r_+-T\right)\right],
\ee
Requiring both $P'(r_{+})=P''(r_{+})=0$, we can obtain
\bea\label{critiPT}
P_c&=&\frac{\pm\sqrt{k^2 \epsilon ^2 \left(9 \epsilon ^2-8\right)}+3k \epsilon ^2-2k}{32 \pi  r_+^2},\no\\
T_c&=&\frac{\pm 3 \sqrt{k^2 \epsilon ^2 \left(9 \epsilon ^2-8\right)}+9 k \epsilon ^2-4 k }{24 \pi  r_+}.
\eea
Substituting them into Eq.(\ref{tem}), we can obtain the interesting relations:
\be\label{eps}
2 r_+^2 \left[k \left(4-9 \epsilon ^2\right)\pm 6 \sqrt{k^2 \epsilon ^2 \left(9 \epsilon ^2-8\right)}\right]=3 Q^2.
\ee

\subsection{$Q=0$ case}

When $Q=0$, the equation of state reduces to
\be
P=\frac{\sqrt{r_+^4 \left(k^2 \left(4-3 \epsilon ^2\right)-8 \pi  k r_+ T \left(3 \epsilon ^2-2\right)+16 \pi ^2 r_+^2 T^2\right)}-k r_+^2+4 \pi  r_+^3 T}{16 \pi  r_+^4}.
\ee
And Eq.(\ref{eps}) turns into
\be
2 r_+^2 \left[k \left(4-9 \epsilon ^2\right)\pm 6 \sqrt{k^2 \epsilon ^2 \left(9 \epsilon ^2-8\right)}\right]=0.
\ee
Excluding the $r_{+}=0$ case, it gives
\be\label{epsq0}
k \left(4-9 \epsilon ^2\right)\pm 6 \sqrt{k^2 \epsilon ^2 \left(9 \epsilon ^2-8\right)}=0.
\ee

\medskip

\textbf{(1). $k=0$}

\medskip

In this case, the entropy and temperature are always positive. The equation of state has a simple form:
\be
P=\frac{T}{2 r_+}.
\ee
And Eq.(\ref{eps}) always holds. Obviously, in this case, no $P-V$ criticality exists.

\medskip

\textbf{(2). $k=1$}

\medskip

In this case, the equation of state turns into
\be
P=\frac{\sqrt{16 \pi ^2 r_+^2 T^2-8 \pi  r_+ T \left(3 \epsilon ^2-2\right)-3 \epsilon ^2+4}+4 \pi  r_+ T-1}{16 \pi  r_+^2}.
\ee
Eq.(\ref{epsq0}) becomes
\be
4-9 \epsilon ^2\pm6 \sqrt{\epsilon ^2 \left(9 \epsilon ^2-8\right)}=0.
\ee
It can be easily checked that the equation with minus sign ``-"  has no real roots for $\epsilon$. And the equation with plus sign ``+"  has the real roots:
\be\label{epsc1}
\epsilon_c=\pm \sqrt{\frac{4}{9}+\frac{8}{9 \sqrt{3}}}\approx \pm 0.9786.
\ee
According to Eq.(\ref{critiPT}), one can obtain
\be
T_c=\frac{1}{2 \sqrt{3} \pi  r_+}, \quad P_c=\frac{2 \sqrt{3}-1}{48 \pi  r_+^2}.
\ee
For any value of $r_{+}$, there is the universal relation:
\be
\frac{P_cr_{+}}{T_c}=\frac{2 \sqrt{3}-1}{8 \sqrt{3}}\approx 0.178.
\ee

\begin{figure}[htp]
\center{
\includegraphics[width=7cm,keepaspectratio]{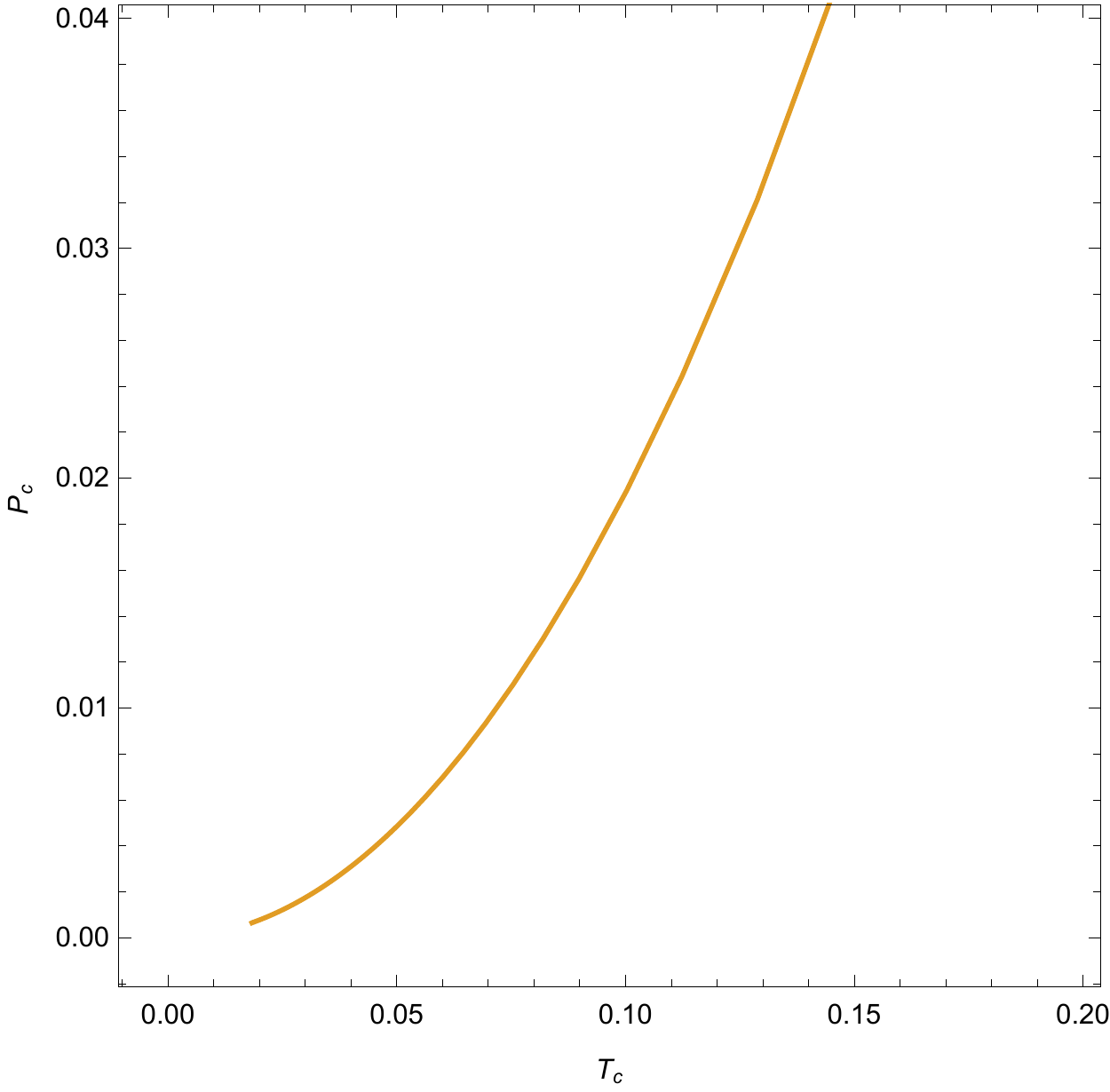}
\caption{The $P_c-T_c$ critical curve of a topological HL black hole for $k=1$. }\label{PTk1}}
\end{figure}

Therefore, in this case it is not a critical point, but a ``critical curve". As is shown in Fig.\ref{PTk1}, any point in this curve is the critical point. This critical behavior is not analogous to those of a VdW system or RN-AdS black hole or other black holes in AdS space, for which there is only one critical point.

\begin{figure}[htp]
\center{
\includegraphics[width=7cm,keepaspectratio]{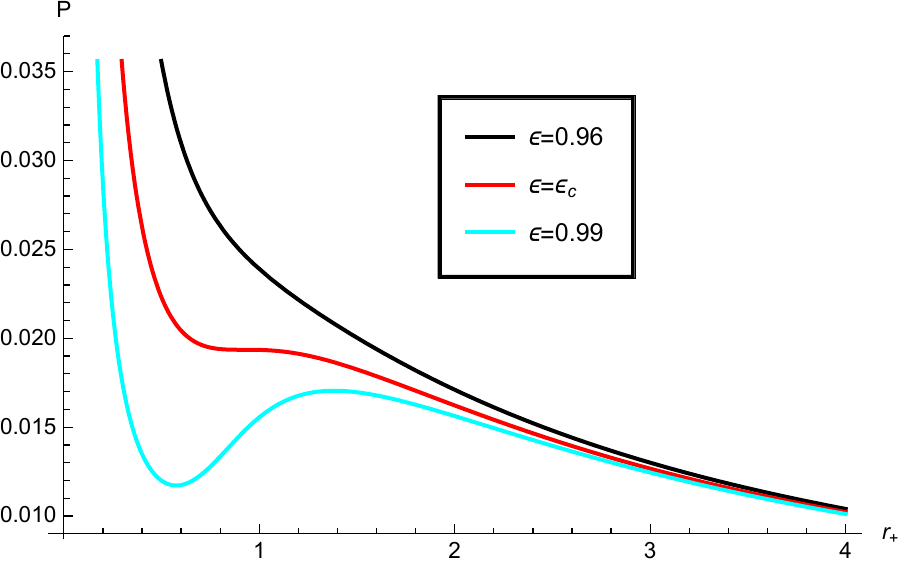}\hspace{0.2cm}
\includegraphics[width=7cm,keepaspectratio]{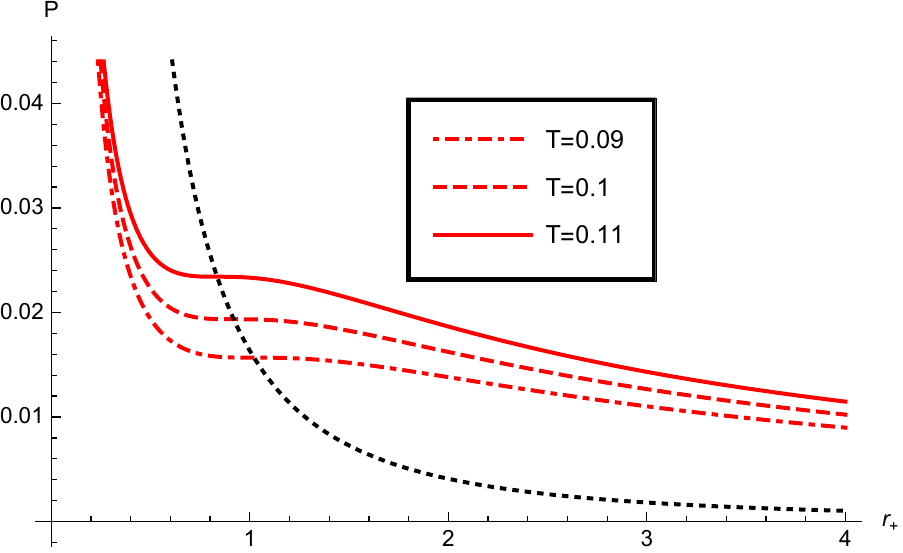}
\caption{The $P-V/r_{+}$ criticality of a topological HL black hole for $k=1$. In the left panel, the temperatures of the three isotherms are all $T=0.1$. In the right panel, the parameter $\epsilon$ is fixed to be $\epsilon_c$. The dashed (black) line describes the position of the critical points.}\label{Prk1}}
\end{figure}

The corresponding $P-r_{+}$ diagrams are depicted in Fig.\ref{Prk1}. The left plot of Fig.\ref{Prk1} shows that the $P-r_{+}$ curves are exactly the same as the $P-v$ curves of a VdW liquid/gas system. However, there is an important difference. The critical behaviors of the uncharged HL black hole are controlled by the parameter $\epsilon$, but not the temperature $T$. The three curves in this plot have the same temperature. The values of $\epsilon$ determine the presence or absence of the $P-r_{+}$ criticality.  When $\epsilon>\epsilon_c$, there is an unstable region with a negative compression coefficient in the $P-r_{+}$ curve, and it should be replaced by a horizontal line determined by Maxwell's equal area law. In this case, the smaller black hole/larger black hole phase transition occurs, which is reminiscent of the liquid/gas phase transition of the VdW system.  When $\epsilon<\epsilon_c$, no phase transition occurs. As is shown in the right plot of Fig.\ref{Prk1}, only if $\epsilon=\epsilon_c$, the isotherms are critical ones regardless of the temperature. The different values of temperature only influence the position of the critical points.

\begin{figure}[htp]
\center{
\includegraphics[width=7cm,keepaspectratio]{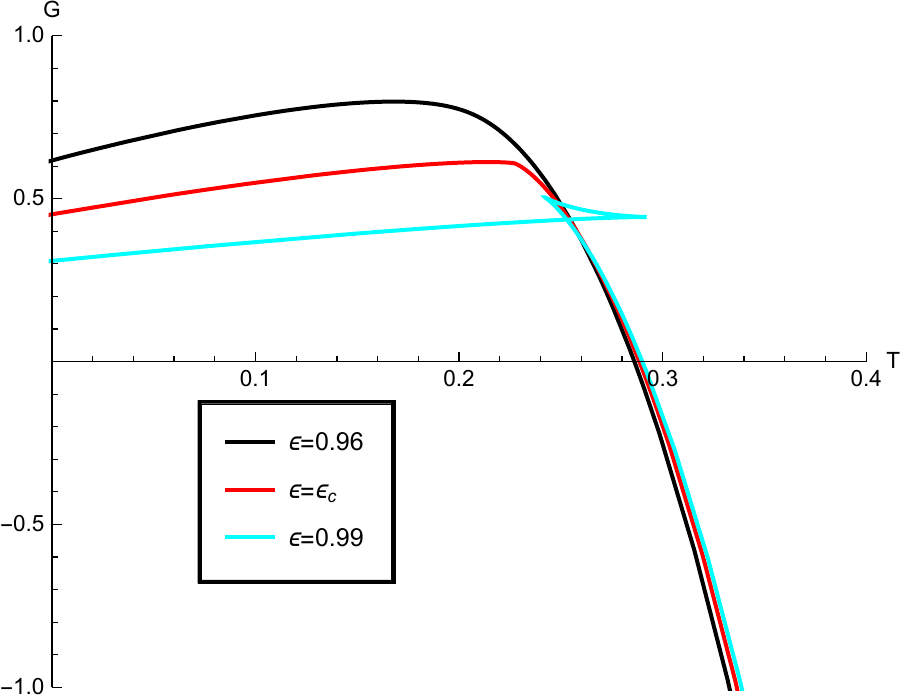}
\caption{The $G-T$ diagram of a topological HL black hole for $k=1$ at constant pressure $P=0.1$. }\label{GTk1}}
\end{figure}

\begin{figure}[htp]
\center{
\includegraphics[width=7cm,keepaspectratio]{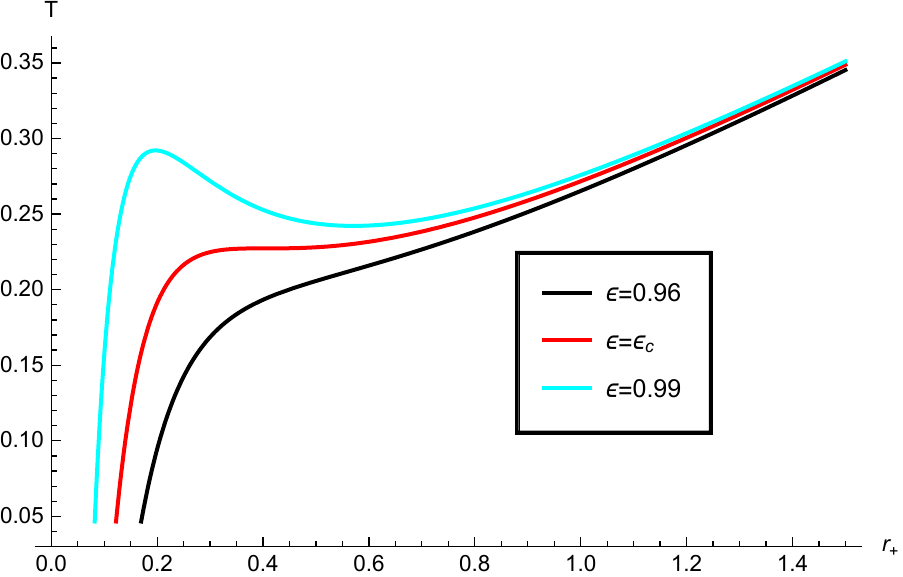}
\caption{The $T-r_{+}$ diagram of a topological HL black hole for $k=1$. We set $P=0.1$. }\label{Trk1}}
\end{figure}

In the left plot of Fig.\ref{Prk1}, one can notice another interesting behavior. When $P>P_c$, the constant pressure line can intersect with the curves at three points, and when $P<P_c$, the constant pressure line can intersect with these curves at five points. These points represent degenerate thermodynamic states with the same pressure and temperature. To ascertain which one of them is thermodynamically preferred, we need to calculate the Gibbs free energy. According to Fig.\ref{GTk1}, when $\epsilon>\epsilon_c$, the $G-T$ diagrams exhibit ``swallow tail" behavior, which is a typical feature of the first-order phase transition. We also find that at lower temperatures the HL black hole with a larger value of $\epsilon$ is more thermodynamically stable and at higher temperatures the HL black hole with a smaller value of $\epsilon$ is more thermodynamically stable. Considering the $T-r_{+}$ relations in Fig.\ref{Trk1}, it means that at a low temperature the smaller HL black hole is more thermodynamically stable and at a high temperature the larger HL black hole is more thermodynamically preferred.

\medskip

\textbf{(3). $k=-1$}

\medskip

In this case, Eq.(\ref{epsq0}) turns into
\be
9 \epsilon ^2-4\pm6 \sqrt{\epsilon ^2 \left(9 \epsilon ^2-8\right)}=0.
\ee
We find that the equation with ``+" sign has no real roots and the equation with ``-" sign has the same real roots as those in Eq.(\ref{epsc1}). However, in this case the critical pressure and the critical temperature are both negative,
\be
T_c=-\frac{1}{2 \sqrt{3} \pi  r_+}, \quad P_c=\frac{1-2 \sqrt{3}}{48 \pi  r_+^2}.
\ee
In other words, there do not exist $P-V$ criticality and phase transition in this case.

\begin{table}[!hbp]
\centering
\begin{tabular}{|c|c|c| }
\hline\hline
 ~ & $k=1$ & $k=-1$ \\
\hline
$k \left(4-9 \epsilon ^2\right)+6 \sqrt{k^2 \epsilon ^2 \left(9 \epsilon ^2-8\right)}>0$ & $\epsilon^2>\frac{1}{27} \left(12+8 \sqrt{3}\right)$ or $\epsilon=0$ & $\epsilon^2\geq8/9$ \\
\hline
$k \left(4-9 \epsilon ^2\right)-6 \sqrt{k^2 \epsilon ^2 \left(9 \epsilon ^2-8\right)}>0$ & $\epsilon=0$ & $\frac{8}{9}\leq \epsilon^2<\frac{1}{27} \left(12+8 \sqrt{3}\right)$ \\
\hline
\end{tabular}
\caption{For $k=\pm 1$, under these conditions there exist critical points.}\label{Tab1}
\end{table}

\subsection{$Q \neq 0$ case}

According to Eq.(\ref{eps}), there should be the following requirement on $\epsilon$:
\be\label{epsp0}
k \left(4-9 \epsilon ^2\right)\pm 6 \sqrt{k^2 \epsilon ^2 \left(9 \epsilon ^2-8\right)}>0.
\ee
And the existence of real roots requires $\epsilon^2\geq8/9$ or $\epsilon=0$.
For $k=0$, the above inequality is clearly violated. Thus, no critical point exists in this case.

For $k=\pm 1$, the constraints on $\epsilon$ are given in Table.\ref{Tab1}. According to Eq.(\ref{critiPT}), one can easily check that for $k=-1$ the critical temperature and the critical pressure  are negative when $\epsilon$ is in the regions in Table.\ref{Tab1}. Therefore, in this case there is also no physically acceptable critical behavior. Below we only focus on the $k=1$ case.

For $\epsilon=0$, Eq.(\ref{eps}) has the solution $r_{+}=\sqrt{\frac{3}{8}}Q$ (we always consider positive electric charges.). However, the corresponding critical pressure and critical temperature are both negative. Therefore, this case should be dropped. Thus, the solely feasible choice is  $\epsilon^2 >\frac{4}{27} \left(3+2 \sqrt{3}\right)\approx0.958$. For the special value $\epsilon=1$, the HL gravity will return back to the GR. Correspondingly, the HL black hole is reduced to the RN-AdS black hole. Therefore, in this case it certainly has $P-V$ criticality.

\begin{figure}[htp]
\center{
\includegraphics[width=7cm,keepaspectratio]{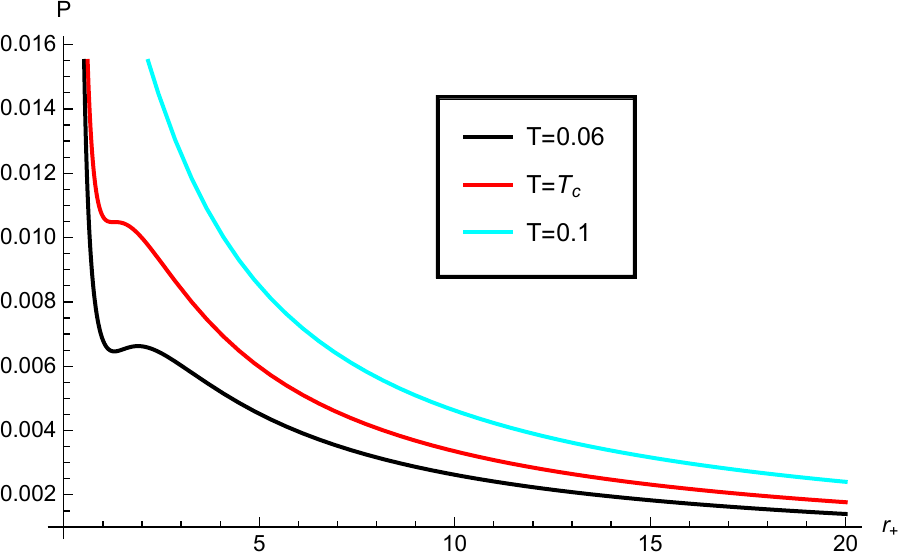}
\caption{The $P-V/r_{+}$ criticality of the charged HL black hole for $k=1$. The critical temperature is $T_c=0.0746$.}\label{prqk1}}
\end{figure}

\begin{figure}[htp]
\center{
\includegraphics[width=7cm,keepaspectratio]{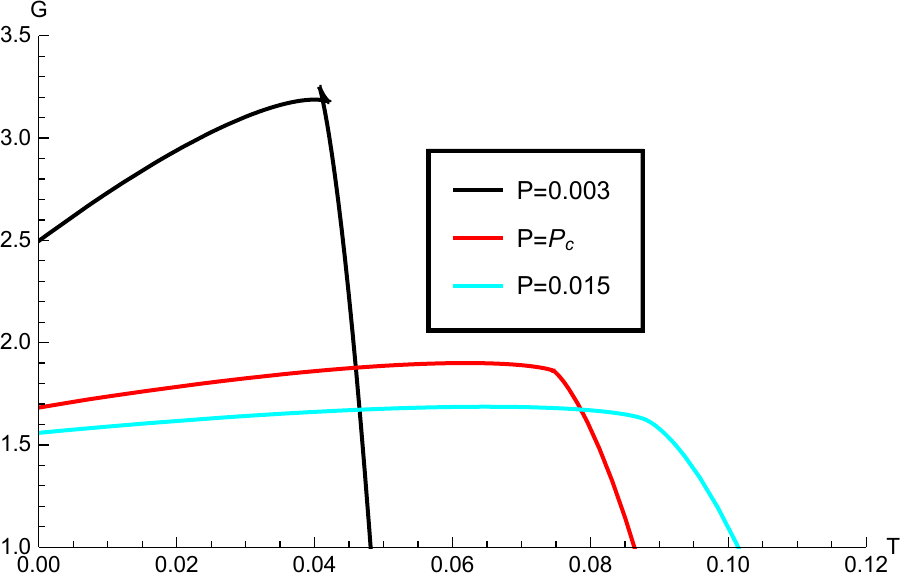}
\caption{The $G-T$ diagram of the charged HL black hole for $k=1$. The critical pressure is $P_c=0.0105$.}\label{GTqk1}}
\end{figure}

When $\epsilon$ situates in the interval $0.958<\epsilon^2 < 1$, we can derive the critical point:
\bea
r_c^2&=&\frac{3 Q^2}{-18 \epsilon ^2+12 \sqrt{\epsilon ^2 \left(9 \epsilon ^2-8\right)}+8},\no\\
P_c&=&\frac{\left(3 \epsilon ^2+\sqrt{\epsilon ^2 \left(9 \epsilon ^2-8\right)}-2\right) \left(-9 \epsilon ^2+6 \sqrt{\epsilon ^2 \left(9 \epsilon ^2-8\right)}+4\right)}{48 \pi  Q^2},\no\\
T_c&=&\frac{(9 \epsilon ^2+3 \sqrt{\epsilon ^2 \left(9 \epsilon ^2-8\right)}-4)\left(-9 \epsilon ^2+6 \sqrt{\epsilon ^2 \left(9 \epsilon ^2-8\right)}+4\right)}{12 \sqrt{6} \pi Q}.
\eea
The universal ratio $\frac{P_cr_c}{T_c}=\frac{3}{32} \left(-3 \epsilon ^2+\sqrt{\epsilon ^2 \left(9 \epsilon ^2-8\right)}+4\right)$, which is independent of the electric charge $Q$. Because the interval for $\epsilon$ is so narrow, we should fine-tune the values of $\epsilon$ and $Q$ to exhibit the $P-r_{+}$ criticality. Fig.\ref{prqk1} and Fig.\ref{GTqk1} show the criticality and phase transition of the HL black holes with $\epsilon=0.985$ and $Q=0.6$. In general, the charged HL black hole for $k=1$ has the similar critical behaviors to that of RN-AdS black hole.

We can also calculate the critical exponents of the charged HL black hole. Introducing the following dimensionless quantities:
\be
t=\d{T}{T_c}-1, \quad \Delta=\d{r_{+}}{r_c}-1, \quad p=\d{P}{P_c}
\ee
and replacing $r_{+},~T$ and $P$ in Eq.(\ref{Pre}) with the new dimensionless parameters $\Delta,~t$ and $p$ and then expanding the equation near the critical point approximately, after chopping some very small numbers one can obtain
\be\label{app}
p=1+At+Bt\Delta+C\Delta^3+O(t^2,t\Delta^2,\Delta^4),
\ee
where $A,~B$ and $C$ are all functions of $\epsilon$ . Eq.(\ref{app}) has the same form as that for the van der Waals system and the RN-AdS black hole\cite{Kubiznak.2012}. Therefore, for this system
the critical exponents should also be $\beta=1/2,~\gamma=1$ and $\delta=3$. In addition, because $C_V=0$, we also have the critical exponent $\alpha=0$.
Obviously, they obey the scaling symmetry like the ordinary thermodynamic systems.

\section{Conclusion and Discussion}
\label{Conclusions}

We considered the HL gravity model with a nonzero $\epsilon$. When the parameter $\epsilon^2$ increases from zero to one, the gravitational theory under consideration changes from the usual HL gravity to general relativity. Restricting $0\leq \epsilon^2\leq 1$, we studied the $P-V$ criticality of topological HL black hole in the extended phase space. We found that the HL black holes with the flat horizon ($k=0)$ and the hyperbolic horizon ($k=-1$) have no $P-V$ criticality and phase transition. For the HL black hole with the spherical horizon ($k=1$), with or without the electric charge, the $P-V$ criticality can appear. In the charged case, the HL black hole exhibits the similar critical behaviors and phase transition to that of a VdW liquid/gas system or RN-AdS black hole.

What is even more interesting is the uncharged HL black hole. In this case, we found some peculiar critical phenomena. First, there is an infinite number of critical points. The critical temperature $T_c$ and the critical pressure $P_c$ both depend on $r_{+}$, while $r_{+}$ is free and is subjected to no constraint. Any value of $r_{+}$ corresponds to the critical position $r_c$. So, we have a ``critical curve", but not only a critical point. Second, for the uncharged HL black hole it is the parameter $\epsilon$ that controls the $P-V$ criticality. This is very different from  the $P-V$ criticality in the VdW liquid/gas system, where the temperature controls the critical behavior. We found that there is a critical $\epsilon=\epsilon_c$, above which the smaller/larger black hole phase transition occurs. Especially, the $P-V$ criticality is completely determined by the parameter $\epsilon$. By that we mean that only if $\epsilon=\epsilon_c$, any temperature is the critical temperature.

We studied the HL black hole with $\lambda=1$. It is of great interest to extend our current study to a slightly more general $\lambda$. Another interesting future study would be to consider the phase transition and thermodynamic stability of the HL black holes in the non-extended phase space.
It is also interesting to consider whether the peculiar $P-V$ criticality we found also exists in other black holes or thermodynamic systems.

Note added: After this work was published in PRD,  the authors learned that the phenomena of critical curve have been found lately\cite{Hennigar.2017a,Hennigar.2017b,Dykaar.2017}.  So this peculiar $P-V$ criticality is of course not found by us for the first time.
However, in our work there are some features that different from previous works.  Our finding occurs for the black hole with spherical horizons in four-dimensional spacetime and the Lagrangian contains only quadratic powers of the curvature.

\appendix

\section{The expression for the event horizon $x_{+}$}

\be
x_{+}=\frac{\sqrt{\frac{12 \sqrt{6} m}{\sqrt{B}}-B-24 k}+\sqrt{B}}{2 \sqrt{6}},
\ee
where
\bea
B&=&\frac{4 \sqrt[3]{2} \left(k^2 \left(8-6 \epsilon ^2\right)+3 q^2\right)}{\sqrt[3]{A}}+2^{2/3} \sqrt[3]{A}-8 k;\\
A&=&-\sqrt{\left(16 k^3 \left(9 \epsilon ^2-8\right)-72 k q^2+27 m^2\right)^2+32 \left(k^2 \left(6 \epsilon ^2-8\right)-3 q^2\right)^3}\no \\
   &~&~~-72 k q^2+27 m^2+144 k^3 \epsilon ^2-128 k^3.
\eea

\acknowledgments
M.-S. M would  like to thank Professor Ren Zhao for illuminating conversations.
This work is supported in part by the National Natural Science Foundation
of China (Grants No.11605107 and No. 11475108) and by the Doctoral Sustentation Fund of Shanxi Datong
University (2011-B-03).

\bibliographystyle{JHEP}
%\bibliography{H:/mms/References/HLgravity}

\end{document}